# THE LENS PARALLAX METHOD: DETERMINING THE REDSHIFTS OF THE FAINT BLUE GALAXIES THROUGH GRAVITATIONAL LENSING

Ramesh Narayan[1] and Matthias Bartelmann[1,2]

[1] Harvard-Smithsonian Center for Astrophysics
60 Garden Street, Cambridge, MA 02138

[2] Max-Planck-Institut für Astrophysik
Karl-Schwarzschild-Strasse 1, D–85740 Garching, FRG

## ABSTRACT

We propose a new technique, which we call the lens parallax method, to determine simultaneously the redshift distribution of the faint blue galaxies and the mass distribution of foreground clusters of galaxies. The method is based on gravitational lensing and uses the fact that the mean angular size of the faint blue galaxies is a well-determined function of surface brightness. The method requires first a calibration of the angular-size vs. surface brightness relation in unlensed blank fields and a determination of the mean redshift of the brightest of the faint blue galaxies. By combining this information with observations of the distorted images of background galaxies in the fields of foreground clusters, we show that it is possible to obtain the mean redshift of the galaxies as a function of their surface brightness. With a sample of about ten moderate redshift rich clusters and using ten bins in surface brightness, a bin-to-bin signal-to-noise ratio of $\sim 3.5$ can be achieved. The method simultaneously allows a determination of the convergence and shear of the lensing cluster as a function of position, and through this the mass distribution of the lens can be obtained. The lens parallax method can be used in conjunction with, and will improve the accuracy of, other previously proposed techniques for mapping clusters.

*Subject headings:* Cosmology: Gravitational Lensing – Galaxies: Clustering – Galaxies: Distances and Redshifts – Methods: Observational

## INTRODUCTION

The effect of gravitational lensing on the images of background galaxies is twofold: the images are magnified, and they are sheared. However, quantities like surface brightness and colors are preserved. The techniques described in this paper are based on these facts. We use a measurement of surface brightness to identify the sub-class to which a particular lensed galaxy belongs. We then compare the observed angular size and elliptic distortion of the galaxy image to what we expect in the absence of lensing and thereby determine the redshift of the galaxy as well as the properties of the lens.

Observations show that there is a well-defined relation between the angular scale length $\theta_s$ and the apparent $B$ magnitude of the faint blue galaxies, that is to say, the scatter around the average angular size $\bar\theta_s(B)$ at a given $B$ is fairly small, $\Delta\theta_s(B) \ll \bar\theta_s(B)$ (Tyson 1994). Equivalently, there is a tight relation between $\bar{\theta}_s$ and the (blue) surface brightness $I$. Gravitational lensing conserves $I$ but multiplies the scale size $\theta_s$ by $\sqrt{\mu}$, where $\mu$ is the magnification induced by the lens. In addition to an overall magnification, the shear in the lens also induces a distortion in the shape of the image.

For source redshifts $z_s \gtrsim 0.5$, the surface brightness $I$ of the faint blue galaxies decreases with increasing $z_s$ (e.g., Koo & Kron 1992, Tyson 1994). At the same time, for a given lens, the lens-induced magnification and shear of the galaxy images increase with increasing $z_s$. Thus, for a given foreground lens such as a cluster of galaxies, the effects of lensing increase monotonically with decreasing $I$. By measuring the sizes and shapes of the faint blue galaxies as functions of $I$ we can thus establish a "redshift ladder" for the galaxies. The ladder can be calibrated by direct spectroscopic measurement of the mean redshift of the brightest faint blue galaxies. Using this, the mean redshift, $\bar{z}_s(I)$, of galaxies at fainter $I$ can then be obtained. This in essence is the *lens parallax* method which we describe in this paper. A virtue of the method is that measurements from many clusters can be combined to reduce the noise in the derived $\bar{z}_s(I)$.

As a by-product, the method also allows a superior reconstruction of the mass distribution of the lensing clusters. Current methods (Kaiser & Squires 1993, Broadhurst, Taylor & Peacock 1994, Seitz & Schneider 1994) suffer because they lack information on the redshifts of the lensed sources. This problem will be eliminated once the relation $\bar{z}_s(I)$ is obtained using our method. Furthermore, many current techniques (e.g. Kaiser & Squires 1993) make use only of the shear-induced distortion of the image and need to carry out an integral transform of the observations to calculate the surface mass density of the lens. In our method, we measure both the shear and magnification of the lens and therefore we obtain direct local measurements of the lens surface density. In this context, our method is similar to that described by Broadhurst et al. (1994).

## MEASUREMENTS

The size and shape of a galaxy image can be quantified by the scale length $\theta_s$ of its surface-brightness profile, the ellipticity $\chi$, and the position angle $\varphi_\chi$ of the elliptic distortion. We adopt the common definition,

$$\chi \equiv \frac{a^2 - b^2}{a^2 + b^2}, \tag{1}$$

where $a$ and $b$ are the semi-major and semi-minor scale lengths, and we write $\chi_1 = \chi \cos(2\varphi_\chi)$, $\chi_2 = \chi \sin(2\varphi_\chi)$. For the surface brightness $I$, we use either the central surface brightness or, more conveniently, the total flux divided by $ab$. Both quantities are invariant under lensing.

The method proposed here requires that $\theta_s$ and $I$ be measured first for a large number of faint blue galaxies in blank fields. These measurements will yield a relation $\bar{\theta}_{s,f}(I)$ between the mean $\bar{\theta}_s$ of the field galaxies and their surface brightness $I$, and also the dispersion in values of $\theta_s$ at a given $I$. This provides the calibration for later measurements of galaxies imaged by clusters. It is also desirable that the ellipticity parameter $\chi$ be measured for the field galaxies so as to obtain an estimate of the range of intrinsic ellipticities.

In addition, at the bright end of $I$, the redshifts of an unbiased sample of faint blue galaxies should be measured in order to obtain the mean redshift $\bar{z}_s(I)$ at this value of $I$. Substantial progress has been made in this direction (Peterson et al. 1986, Broadhurst, Ellis & Shanks 1988, Metcalfe et al. 1989, Colless et al. 1990, Cowie et al. 1990, Guhathakurta, Tyson & Majewski 1990, Lilly, Cowie & Gardner 1991, Dressler & Gunn 1992, Koo & Kron 1992), and it is possible that the information available now is already sufficient for the method we propose.

Once these calibrations are done, a number of rich clusters of galaxies at redshifts $z_d \gtrsim 0.3$ should be chosen. Each cluster field should be divided into a number of patches, where each patch is large enough to contain many galaxies, but small enough that the lensing properties of the cluster do not vary significantly within a patch. We will quantify the patch size later. Within each patch, the distorted images of the faint blue galaxies should be used to determine the local mean values of $\bar{\theta}_{s,c}$, $\bar{\chi}_c$ and $\bar{\varphi}_c$ as functions of $I$. By comparing the local $\bar{\theta}_{s,c}(I)$ with the $\bar{\theta}_{s,f}(I)$ of field galaxies, the magnification of the galaxies $\mu(I)$ can be inferred as a function of surface brightness $I$. Similarly, from the local values of $\bar{\chi}_c(I)$, the shear components of the lens can be determined as functions of $I$. As we explain below, this information then allows us to derive both the redshift distribution of the galaxies and the mass density of the lens.

## BASICS OF LENSING

We briefly review here the relevant properties of gravitational lensing (see Blandford & Narayan 1992, Schneider, Ehlers & Falco 1992 for more details.) The local properties of a lens are described by the convergence $\kappa$, the shear $\gamma$, and the position angle $\varphi_\gamma$ of the shear, or equivalently by $\kappa$, $\gamma_1 = \gamma(\cos 2\varphi_\gamma)$, and $\gamma_2 = \gamma \sin(2\varphi_\gamma)$. The convergence $\kappa$ is defined by

$$\kappa(z_d, z_s) \equiv \frac{\Sigma}{\Sigma_{cr}(z_d, z_s)} , \quad \Sigma_{cr}(z_d, z_s) \equiv \frac{c^2}{4\pi G} \frac{D_s}{D_d D_{ds}} , \qquad (2)$$

and is the surface mass density $\Sigma$ of the lens expressed in units of the critical density $\Sigma_{cr}$. The quantities $D_{d,s,ds}$ are angular-diameter distances from the observer to the lens and to the source, and from the lens to the source, respectively. Let us write the convergence $\kappa(z_d, z_s)$ and the shear components $\gamma_i(z_d, z_s)$ as

$$\kappa(z_d, z_s) = \kappa_\infty(z_d) f(z_d, z_s) , \quad \gamma_i(z_d, z_s) = \gamma_{i,\infty}(z_d) f(z_d, z_s) , \quad f(z_d, z_s) \equiv \frac{D_{ds}}{D_s} , \quad (3)$$

where $\kappa_\infty(z_d)$ and $\gamma_\infty(z_d)$ refer to a source at infinity. The advantage of this split is that the properties of the lens are described just by $\kappa_\infty(z_d)$ and $\gamma_\infty(z_d)$, while the variations induced by changes in the source redshift are described by the factor $f(z_d, z_s)$. Fig. 1 illustrates the behavior of $f(z_d, z_s)$ as a function of $z_s$ for four values of $z_d$.

Given measurements of $\bar{\theta}_{s,f}(I)$ and $\bar{\theta}_{s,c}(I)$ in a given patch at a given surface brightness $I$, the magnification $\mu(I)$ can be calculated by

$$\mu(I) = \left[\frac{\bar{\theta}_{s,c}(I)}{\bar{\theta}_{s,f}(I)}\right]^2 , \qquad (4)$$

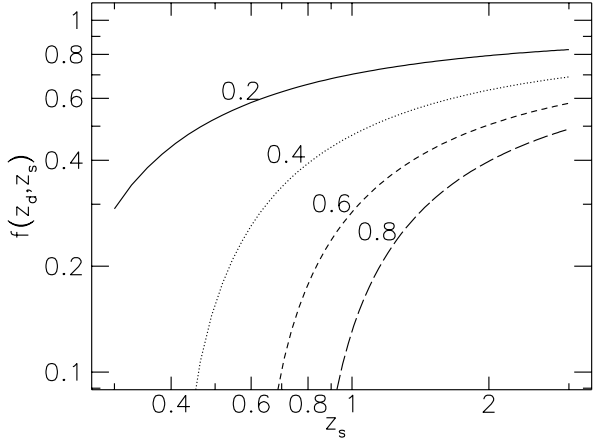

Fig. 1 — The function $f(z_d, z_s)$ in eq. (3) for four different values of $z_d$, as indicated.

while $\mu(I)$ itself is related to the corresponding $\kappa(I)$ and $\gamma(I)$ by

$$\mu(I) = \frac{1}{[1 - \kappa(I)]^2 - \gamma(I)^2} \ . \tag{5}$$

The ellipticity parameters $\chi_1$, $\chi_2$ of a galaxy image are related to the intrinsic ellipticity of the source via a transformation which depends on $\kappa$ and $\gamma$. Since the orientation of the intrinsic ellipticities of the sources is random, it is possible to calculate from the measured $\chi_1$, $\chi_2$ values of galaxies of a given surface brightness, two quantities, $\delta_1(I)$, $\delta_2(I)$, which describe the local distorting effect of the lens corresponding to that $I$ (Kochanek 1990, Miralda-Escudé 1991, Schneider & Seitz 1994). The quantities $\delta_1(I)$, $\delta_2(I)$ are related to $\kappa(I)$ and $\gamma(I)$ via

$$\delta_{1,2}(I) = \frac{2[1 - \kappa(I)]\gamma_{1,2}(I)}{[1 - \kappa(I)]^2 + \gamma(I)^2} \ . \tag{6}$$

Equations (5) and (6) can in general be inverted to estimate the convergence $\kappa(I)$ and shear $\gamma(I)$ as functions of $I$ in each patch of a lens field. In the limit of weak lensing ($\kappa, \gamma \ll 1$), these equations simplify to

$$\mu(I) = 1 + 2\kappa(I), \qquad \delta_{1,2}(I) = 2\gamma_{1,2}(I). \tag{7}$$

Therefore, in this limit, the change of the angular sizes of sources relative to an unlensed field directly measures the convergence, while the induced ellipticity in the images measures the shear.

## DESCRIPTION OF THE METHOD

Let us bin the galaxies into a number of surface brightness bins, $I_i$. For each patch in the lens field, and each brightness bin $I_i$, we obtain via eqs. (5) and (6) the convergence $\kappa(I_i)$ and the shear components, $\gamma_1(I_i)$, $\gamma_2(I_i)$. Compare now the results in two separate brightness bins, $I_i$ and $I_j$, within a given patch. From eq. (3) we see that

$$\frac{\kappa(I_j)}{\kappa(I_i)} = \frac{\gamma_1(I_j)}{\gamma_1(I_i)} = \frac{\gamma_2(I_j)}{\gamma_2(I_i)} = \frac{f[z_d, \bar{z}_s(I_j)]}{f[z_d, \bar{z}_s(I_i)]} \equiv R_{ij}(z_d). \tag{8}$$

The quantity $R_{ij}(z_d)$ is a kind of redshift ladder, which for a fixed $z_d$ is a function only of the mean redshifts, $\bar{z}_s(I_i)$ and $\bar{z}_s(I_j)$, in the two brightness bins, $I_i$ and $I_j$. Equation (8) shows that, in each patch of the lens field, we have three independent estimates of $R_{ij}$. Moreover, $R_{ij}$ should be the same for all patches in the lens. Thus, from $N_{\text{patch}}$ patches, we obtain $3N_{\text{patch}}$ independent determinations of $R_{ij}$.

Let us assume that we know the mean redshift $\bar{z}_s(I_1)$ of the galaxies in the first (i.e. brightest) surface brightness bin. Then, assuming a cosmographic model (e.g. $\Omega_0 = 1$, $\Lambda = 0$), we can directly obtain from $R_{ij}(z_d)$ the mean redshift $\bar{z}_s(I_i)$ of all the other brightness bins. In any given patch the accuracy will be very poor. However, we can average over all the patches in a lens and also average over different clusters (which can be at different redshifts). Thus, the accuracy increases as more data become available. The power of the method lies in the fact that every new measurement can be combined with all previous measurements to improve the overall accuracy of the redshift determinations.

Finally, once the redshift ladder has been used to obtain all the $\bar{z}_s(I_i)$, then eq. (3) gives the lens convergence $\kappa_\infty$ and shear components $\gamma_{1,\infty}$, $\gamma_{2,\infty}$ in each patch of each lens. From this surface density maps of all the lensing clusters are obtained directly (e.g. eq. 2).

## ESTIMATE OF SIGNAL-TO-NOISE RATIO

In order to estimate the signal-to-noise ratio that we can expect with the proposed method, we make the following simplifying assumptions. (1) We take the number density of faint blue galaxies to be $\simeq 3 \times 10^5$ deg$^{-2}$ = 83 arcmin$^{-2}$, which corresponds to a limiting magnitude of $B \simeq 27$ (Koo 1986, Guhathakurta, Tyson & Majewski 1990, Gardner, Cowie & Wainscoat 1993, Tyson 1994). We assume that the area $(\Delta\vartheta)^2$ of the observed patch is chosen such that the number of galaxies in the patch, $\simeq 83\,\Delta\vartheta^2$, is much larger than unity. We also assume that we have 10 surface brightness bins so that the number of galaxies per patch per bin is $N \simeq 8.3\,\Delta\vartheta^2$. (2) We take the intrinsic scatter in the scale lengths $\theta_s$ of galaxies with a given $I$ to be $\pm 20\%$ (Fig. 5 in Tyson 1994). The local $\kappa(I)$ corresponding to a given bin of $I$ in a selected patch in the lens field can therefore be determined to an accuracy of $\simeq 0.2/\sqrt{N}$. (3) We take the rms intrinsic ellipticity of the faint blue galaxies to be $\sqrt{\langle\chi^2\rangle} \simeq 0.3$ (Tyson & Seitzer 1988, Miralda-Escudé 1991). Since in the weak-lensing regime $\langle\chi\rangle \simeq \delta \simeq 2\gamma$, we assume an uncertainty in the local determination of shear of $\simeq 0.15/\sqrt{N}$. (4) We model a "generic" cluster as an isothermal sphere with an Einstein radius of $\simeq 30''$ for a source at infinity. Thus, $\kappa_\infty = \gamma_\infty = 0.25/\vartheta$, where $\vartheta$ is the angular separation in arc minutes from the cluster center. (5) We assume that the faint blue galaxies are distributed in the redshift range $z_s \simeq 1 - 3$. If we take a cluster redshift of $z_d \simeq 0.5$, the function $f(z_d, z_s)$ varies by about a factor of two across this range of $z_s$ (see Fig. 1). With 10 bins, the fractional change in $\kappa$ and $\gamma$ from one bin to the next is then $(\Delta\kappa/\kappa_\infty) \simeq (\Delta\gamma/\gamma_\infty) \simeq 0.05$.

According to assumptions (2) and (3), the noise in the determinations of $\kappa(I_i)$ and $\gamma(I_i)$ in each brightness bin of a patch are $\Delta\kappa \simeq (0.2/\sqrt{8.3})\Delta\vartheta \simeq 0.069/\Delta\vartheta$ and $\Delta\gamma \simeq 0.052/\Delta\vartheta$. By assumptions (4) and (5), the actual change in $\kappa$ and $\gamma$ from one bin to the next is $\simeq 0.05\kappa_\infty \simeq 0.013/\vartheta$. Thus the effective signal-to-noise ratio with which the differences in $\kappa$ and $\gamma$ between two neighboring bins in a single patch can be determined is

$$\left(\frac{S}{N}\right) \simeq \sqrt{\left(\frac{0.013/\vartheta}{0.069/\Delta\vartheta}\right)^2 + \left(\frac{0.013/\vartheta}{0.052/\Delta\vartheta}\right)^2} \simeq 0.30\,\frac{\Delta\vartheta}{\vartheta}\,. \tag{9}$$

To determine the redshift distribution $\bar{z}_{\rm s}(I)$ of the galaxies, we average over all the patches in a single lens as well as over different lenses. Let us therefore integrate $(S/N)^2$ over an area extending from the critical radius at $\vartheta \simeq 0.5$ out to $\vartheta \simeq 5$, and let us assume that we have data on $N_{\rm cl}$ clusters. The net signal-to-noise ratio is then

$$\left(\frac{S}{N}\right)^2 \simeq N_{\rm cl} \int_{0.5}^{5} \left(0.3\,\frac{\Delta\vartheta}{\vartheta}\right)^2 \frac{2\pi\vartheta d\vartheta}{(\Delta\vartheta)^2} \simeq 1.3\,N_{\rm cl}\,. \tag{10}$$

With data on say 10 clusters, we see that we can achieve a respectable signal-to-noise ratio of $(S/N) \simeq 3.5$. What this means is that we will be able to determine the change in $\kappa(I_i)$ and $\gamma(I_i)$ from one brightness bin to the next, and therefore the change in mean redshift $\bar{z}_{\rm s}(I_i)$, with an error of only $\sim 30\%$ of the difference. Of course, if we look at the accuracy across the entire 10 bins, i.e. from the brightest to the faintest surface brightness galaxies, the fractional error will be very much smaller.

Consider next the accuracy that we can expect in the reconstruction of the lens surface mass density $\Sigma$. Now we need to consider each patch of each lens separately since we do not have the option of averaging over different patches. It is therefore critical to choose the patch size appropriately. Assuming we have determined $\bar{z}_{\rm s}(I)$ of all the brightness bins, in this stage of the calculations we can use all the galaxies in a given patch to determine the local lens parameters, $\kappa_\infty$, $\gamma_{1,\infty}$, $\gamma_{2,\infty}$. By assumptions (2) and (3), the accuracies with which the local $\kappa_\infty$ and $\gamma_\infty$ of a patch can be determined are $\simeq 0.022/\Delta\vartheta$ and $\simeq 0.016/\Delta\vartheta$, respectively. Thus, the effective signal-to-noise is

$$\left(\frac{S}{N}\right)_\kappa \simeq \frac{\kappa_\infty \Delta\vartheta}{0.022} \simeq 11\,\frac{\Delta\vartheta}{\vartheta}\,, \quad \left(\frac{S}{N}\right)_\gamma \simeq \frac{\gamma_\infty \Delta\vartheta}{0.016} \simeq 16\,\frac{\Delta\vartheta}{\vartheta}\,. \tag{11}$$

In the simplest scheme, we would just take the derived $\kappa_\infty$ in each patch and calculate the local surface density $\Sigma$ through eq. (2). In this case, the signal-to-noise ratio is just $11\Delta\vartheta/\vartheta$. However, we note that $\kappa_\infty$ and $\gamma_\infty$ are not independent, but are related through a common scalar potential (e.g. Kaiser & Squires 1993). Therefore, a more sophisticated algorithm would combine the estimates of $\kappa_\infty$ and $\gamma_\infty$ across the whole lens field and would thereby obtain a global reconstruction of the surface density distribution. We do not describe the details of such an algorithm here, but note that the effective signal-to-noise ratio in any such scheme is likely to combine (in quadrature) the individual $S/N$ in $\kappa$ and $\gamma$. Thus, the $(S/N)$ we can expect in the reconstructed $\Sigma$ is

$$\left(\frac{S}{N}\right)_\Sigma \simeq 19\,\frac{\Delta\vartheta}{\vartheta}\,. \tag{12}$$

This shows that if our goal is to achieve a certain target signal-to-noise ratio $(S/N)_{\text{target}}$ for the reconstruction of $\Sigma$, then we must select the angular scale of the patches to be

$$\frac{\Delta\vartheta}{\vartheta} \simeq \frac{(S/N)_{\text{target}}}{19} \ . \qquad (13)$$

The angular resolution that we can achieve is maximum near the center of the lens and decreases farther out.

Note that all the estimates in this section have been derived under the assumption of weak lensing. In the strong-lensing regime, especially close to critical curves, the real accuracy that can be achieved is likely to be better than our estimate because of the strong dependence of the location of the critical curve on $z_s$. Also, we have assumed that the lens is a smooth singular isothermal sphere. Actual clusters probably have significant substructure and this will lead to a stronger signal and therefore an improved signal-to-noise ratio.

## DISCUSSION

We propose to call the redshift ladder technique described in this paper the *lens parallax* method. Although the technique does not employ true trigonometric parallaxes, it is nevertheless a purely geometrical method. In essence, the lens produces a differential bending of light rays described by the three local parameters, $\kappa_\infty$, $\gamma_{1,\infty}$ and $\gamma_{2,\infty}$. The ray deflections distort the images of background sources in such a way that the magnitude of the distortions depends through simple geometry on the distance of the source behind the lens (eq. 3). We think the analogy with trigonometric parallax is rather close.

There are two key ideas in this paper. First, we show how the lens parallax method can be used to determine the redshifts of faint galaxies as a function of surface brightness. Second, we suggest that by using field galaxies as calibrators we can measure not only the local shear due to a lens, as has been considered in the past, but also the local magnification. These two ideas work best when combined, but they are essentially independent of each other.

The lens parallax method, for instance, can in principle be applied using only information on shear. This is because the quantity $R_{ij}$ in eq. (8) can be estimated from either the convergence or the shear. The signal-to-noise will of course improve if both $\kappa$ and $\gamma$ are used, but in principle $\gamma$ alone will suffice.

We must emphasize that the lens parallax method does not make any assumptions about the properties of distant galaxies or their evolution. All we require is that there should be a reasonably tight relation between surface brightness and redshift, for which there is already strong evidence (Tyson 1994), and that the galaxy populations in different directions (i.e. towards different cluster fields or blank fields) are essentially the same. Thus the method is quite model-independent.

The idea of using field galaxies as calibrators to determine the local magnification $\mu$ of the lens has been proposed in a different form by Broadhurst et al. (1994). They suggest calibrating the galaxy counts in blank fields and determining how the counts vary in a lensed field. We believe our method is simpler and less model-dependent. We also suspect that the signal-to-noise ratio achievable with our method is superior but this needs to be tested with simulations.

As emphasized by Broadhurst et al. (1994), a measurement of the convergence $\kappa$ directly gives the local surface density $\Sigma$ of the lens, whereas in the methods developed so far which are based solely on shear (Kaiser & Squires 1993, Seitz & Schneider 1994) one needs to convert a map of $\gamma$ into a map of $\Sigma$ via an integral transformation. The transform itself is fairly complicated, but in addition the final answer is uncertain up to an overall additive uniform mass density (Schneider & Seitz 1994). Our technique is more direct and does not suffer from the mass ambiguity.

The methods described here can be extended in several ways. First, instead of using only surface brightness to characterize the galaxy population, we could also include other properties that are invariant to lensing, e.g. colors (assuming there is no reddening due to the lenses). Suppose for instance that the dispersion of redshifts per bin becomes smaller when the galaxies are characterized by both surface brightness and color rather than brightness alone. It may then be possible to obtain the mean redshift of the galaxies as a function of both brightness and color wihtout much loss in the signal-to-noise ratio. We will thus achieve a more detailed description of the evolution of the faint blue galaxies. Another possibility is that by comparing the results from lenses at different redshifts $z_d$, we can test for the internal consistency of the independent redshift ladders obtained from each cluster. In principle, this will allow us to constrain the model of the universe. Gravitational lenses are particularly effective at distinguishing models with a large cosmological constant $\Lambda$ (Turner 1990, Kochanek 1993). The lens parallax method may therefore provide a purely geometrical technique for constraining the value of $\Lambda$. Finally, we can test whether or not the entire lens effect is due to the lensing cluster as assumed in our analysis. Under this assumption, the shear angle $\varphi_\gamma(I)$ in a given patch of the lens should be independent of $I$. If the data do not show this, then we could infer that there are significant additional shear contributions at redshifts $z > z_d$ which introduce differential effects as a function of source redshift. This could potentially be a powerful diagnostic of structure at high redshift.

The role of systematic effects in the lens parallax method is unclear at the moment. The method requires that we be able to measure the surface brightnesses, scale lengths and ellipticities of very faint galaxies with reasonable accuracy. The effect of atmospheric seeing may be particularly serious, and it is possible that the method may require space-based observations in order to avoid errors due to variable seeing. One point to note, however, is that a wrong estimate of the seeing leads to errors of opposite signs in $\kappa$ and $\gamma$. It is thus possible that the method may be fairly insensitive to moderate errors in the seeing estimate.

This work was supported in part by NSF grant AST 9109525.

## REFERENCES

Blandford, R.D., & Narayan, R. 1992, ARA&A, 30, 311
Broadhurst, T.J., Ellis, R.S., & Shanks, T. 1988, MNRAS, 235, 827
Broadhurst, T.J., Taylor, A.N., & Peacock, J.A. 1994, ApJ, in press
Colless, M., Ellis, R.S., Taylor, K., & Hook, R.N. 1990, MNRAS, 244, 408

Cowie, L.L., Gardner, J.P., Lilly, S.J., & McLean, I. 1990, ApJ, 360, L1
Dressler, A., & Gunn, J.E. 1992, ApJS, 78, 1
Gardner, J.P., Cowie, L.L., & Wainscoat, R.J. 1993, ApJ, 415, L9
Guhathakurta, P., Tyson, J.A., & Majewski, S.R. 1990, ApJ, 357, L9
Kaiser, N., & Squires, G. 1993, ApJ, 404, 441
Kochanek, C.S. 1990, MNRAS, 247, 135
Kochanek, C.S. 1993, ApJ, 419, 12
Koo, D.C. 1986, ApJ, 311, 651
Koo, D.C., & Kron, R.G. 1992, ARA&A, 30, 613
Lilly, S.J., Cowie, L.L., & Gardner, J.P. 1991, ApJ, 369, 79
Metcalfe, N., Fong, R., Shanks, T., & Kilkenny, D. 1989, MNRAS, 236, 207
Miralda-Escudé, J. 1991, ApJ, 370, 1
Peterson, B.A., Ellis, R.S., Efstathiou, G., Shanks, T., Bean, A.J., Fong, R., & Zen-Long, Z. 1986, MNRAS, 221, 233
Schneider, P., Ehlers, J., & Falco, E.E. 1992, *Gravitational Lenses* (Heidelberg: Springer)
Schneider, P., & Seitz, C. 1994, A&A, submitted
Seitz, C., & Schneider, P. 1994, A&A, submitted
Turner, E.L. 1990, ApJ, 365, 43
Tyson, J.A. 1988, AJ, 96, 1
Tyson, J.A., & Seitzer, P. 1988, ApJ, 335, 552
Tyson, J.A., Valdes, F., & Wenk, R.A. 1990, ApJ, 349, 1
Tyson, J.A. 1994, in: *Cosmology and Large-Scale Structure*, Proceedings Les Houches Summer School, August 1993, ed. R. Schaeffer (Elsevier)